\documentclass[]{elsart}
\usepackage{graphics}
\usepackage{latexsym}

\begin{document}
\begin{frontmatter}

\title{
Scaling Properties of Pion Production
}

\author{T.Peitzmann\thanksref{email}} 
\address{University of M{\"u}nster, D-48149 M{\"u}nster, Germany}
\thanks[email]{email: peitzmann@ikp.uni-muenster.de}
\begin{abstract}
	The scaling of pion multiplicities with 
	system size in proton-nucleus and nucleus-nucleus reactions at 200 
	$A$ GeV is investigated. 
	A nuclear geometry calculation of incoherent multiple scattering 
	including effects of deceleration of the participating particles is 
	presented. Pion multiplicities at high transverse momenta 
	agree relatively well with an overall 
	scaling with the number of binary collisions. The deviations from 
	this gross scaling for p+A and S+Au reactions at 200$A$GeV are 
	compared to model predictions and the 
	implications are discussed.
	Within this picture no universal scaling law for 
	high $p_{T}$ particle production is 
	obtained. In addition to the conceptual problems of 
	such an incoherent model, it is shown that the implied kinematic 
	relation between a 
	possible $p_{T}$-broadening and deceleration 
	leads to contradictions. 
	The scaling of pion production at intermediate $p_{T}$ with 
	system size does not 
	indicate a multiple scattering enhancement.
\end{abstract}

\end{frontmatter}
\section{Introduction}
Hadron production at high transverse momenta in very high energy collisions 
is thought to be described by perturbative QCD. The transverse 
momentum and energy dependence of the cross section is however not 
consistent with simple qq scattering processes. Higher order 
contributions and 
scaling viola\-tions have to be included. At somewhat lower 
energies ($\sqrt{s} \leq 20 \, \mathrm{GeV}$) it is still unclear 
whether meson degrees of freedom contribute to high $p_{T}$ scattering 
via the \emph{constituent interchange mechanism} (for reviews see e.g.
\cite{rmp:owens:87,prp:geist:90}).

The interest in high $p_{T}$ production has been renewed by the 
observation of the so-called \emph{Cronin-effect} in p+A collisions at 
Fermilab \cite{prd:antreasyan:79}. Here it was seen that the target 
mass dependence of particle production when parameterized as 
$A^{\alpha}$ reveals a power larger than one for $p_{T} > 2 \, 
\mathrm{GeV}/c$, an effect, which was later interpreted as 
the result of multiple 
scattering of the incident 
partons \cite{plb:krzywicki:cronin:79,zpc:lev:cronin:83}.

In nucleus-nucleus collisions there is still a 
strong enhancement of particle production at high $p_{T}$. It has 
been proposed to parameterize the nuclear dependence as 
\cite{jpg:schmidt:rhic:93}:   
\begin{equation}
	E \frac{d^{3}\sigma}{dp^{3}}(AB) = (A \cdot B)^{\alpha(p_{T})}
	E \frac{d^{3}\sigma}{dp^{3}}(pp).
	\label{eq:cronin1}
\end{equation}
For minimum bias nucleus-nucleus collisions this parameterization yields 
exponents at high $p_{T}$ which are comparable to those in p+A, 
however, for heavy nuclei 
the statistical significance was so far limited. Recently high 
precision neutral pion data from 200$A$GeV S+Au collisions have become 
available \cite{misc:albrecht:pi0:98}, which corroborate the above 
result. It is also emphasized in \cite{misc:albrecht:pi0:98} that this 
enhancement must not necessarily be an isolated feature at high 
$p_{T}$, but that differences of the shape of the momentum spectra 
over the whole $p_{T}$ range are important.

Still the similarity of the exponents may be accidental, as a lot of 
information from the heavy ion reaction is washed out due to the 
impact parameter averaging. In a na\"{\i}ve picture, the total cross 
section in nucleus-nucleus collisions should scale like 
$(R_{A}+R_{B})^{2} \propto (A^{1/3} + B^{1/3})^{2}$, 
while the yield of produced particles for a given nuclear collision
should depend mainly on the apparent thicknesses like $A^{1/3} \cdot 
B^{1/3}$. For proton-nucleus this reduces to a simple dependence, 
because $A \ll B$. This is not generally true for a heavy-ion 
collisions ($A \neq B$), 
with an exception of the special case where every possible 
binary collision contributes the same amount to the total cross 
section without further modification. Then the cross section must 
be $AB \sigma_{pp}$.In case of an anomalous nuclear 
enhancement as observed in \cite{prd:antreasyan:79} such a simple 
relation does not hold.

I will try to establish a way of investigating scaling properties 
which can be applied to centrality selected heavy-ion data. Special 
attention will be given to high $p_{T}$ production. There it is 
reasonable to assume a scaling with the number of binary collisions 
as a baseline estimate. Modifications due to initial state multiple 
scattering should scale with the thickness of the colliding nuclei. 
The rms-thickness properly averaged over all possible nucleon-nucleon 
collisions will be used here. If the nuclear enhancement is caused by 
initial state multiple scattering, one expects the particle yield per 
binary collision to be a continuous and monotonous function of this 
thickness parameter.

For the intermediate $p_{T}$ range there have been attempts 
\cite{qm96:leonidov}
to describe the change in slope of the spectra for heavier systems as 
a $p_{T}$-broadening due to initial state multiple scattering. There 
only the shapes of the spectra have been compared, no attempt was made 
to predict the absolute scale of particle production. As such a model 
must also account for changes in the multiplicity of particles, it is 
of interest to simultaneously compare the concepts of multiple 
scattering also to data at lower $p_{T}$.

In this paper I use a simple simulation of incoherent 
multiple scattering in the initial state to predict possible 
scaling laws for particle production. 
One has to note, however, that the assumption of \emph{incoherence} 
and the decomposition of the reaction into subsequent binary 
collisions is very na{\"{\i}}ve and most likely not fulfilled.

I will calculate a certain transverse rapidity kick from the 
effective thickness with the help of a hydrodynamically inspired 
formula. This recipe may be questionable, but is used merely to study 
the change in shape of the spectra -- which it should allow -- 
and does not imply any physical significance of the hydrodynamic 
origin of the formula.
In addition, a possible energy degradation is calculated, and the 
altered $\sqrt{s}$ used in the following collisions. This relies 
heavily on the incoherence of the collisions and is in this respect 
very similar to a na{\"{\i}}ve cascade model. It is actually not 
expected to be applicable in high energy reactions.

Also, the transverse rapidity kick is only relevant for the produced 
particles. The number of collisions the nucleons suffer is calculated 
by assuming that they continue to travel on a straight line. It is further 
assumed that both the rapidity kick and the longitudinal rapidity loss 
are independent of the $\sqrt{s}$ of the collision in question. This 
will obviously not be true once $\sqrt{s}$ gets very small. This will 
be neglected because the concept of energy degradation has its 
problems, and we will only make limited use of it in this paper. For 
similar reasons the effects of varying ``source rapidity'' 
depending on the amount of multiple scattering will be neglected. 

The predictions will be 
compared to pion multiplicities in proton-nucleus collisions at 200~GeV 
measured at a pseudorapidity $\eta = 3.26$ 
\cite{prd:antreasyan:79} and S+Au data for different impact parameters at 
200~$A$~GeV in the rapidity range 
$2.1 \le y \le 2.9$ \cite{misc:albrecht:pi0:98}. 
Such comparisons 
might help in understanding the limitations of incoherent 
(``random-walk'') multiple scattering for the description of particle production 
in nucleus-nucleus collisions.
\section{The Model}
The model calculations use the following recipe:
For a given reaction system nucleons are randomly distributed in the nuclei 
according to Woods-Saxon distributions: 
\begin{displaymath}
	\rho(r) \propto \frac{1}{1 + \exp\left[ (r - r_{0})/0.54 \right]}
\end{displaymath}
using $r_{0} = 1.16 A^{1/3} - 1.35 A^{-1/3}$.
It is taken care that the 
hard cores of the nucleons ($R_{hc} = 0.6 \, \mathrm{fm}$) do not 
overlap. The two nuclei collide at a given impact parameter. A 
projectile nucleon is supposed to interact with a target nucleon, if 
their minimum distance on straight line trajectories is less than 
their interaction distance ($d_{int} = 1.025 \, 
\mathrm{fm}$).\footnote{Mostly nucleon-nucleon 
interactions are presumed, 
but it should not really matter whether the nucleon or the 
parton picture determines the underlying dynamics. One of the 
assumptions of the model is, however, that all individual binary 
collisions in the nuclei are incoherent.}
The calculations are repeated for a large number of random configurations 
of nucleons within the nuclei.

From this model the following variables are calculated:
\begin{itemize}
	\item The average number of binary collisions $N_{coll}$, 
	which may happen in a nuclear reaction. This is 
	the main variable controlling particle multiplicities at high momenta 
	in a model of independent production.
	
	\item The reduction in particle multiplicity due to the deceleration 
	of parti\-ci\-pant nucleons. It is assumed that a particle suffers a 
	constant rapidity shift $\delta y$ per collision. The number 
	of collisions a nucleon may 
	undergo before a given collision which is responsible for the 
	production of the particle of interest is calculated. I will call this the 
	\emph{number of prescatterings} $N_{pre}$. It determines the total 
	rapidity loss and thereby the center-of-mass energy of the collision 
	in question $\sqrt{s_{1}}$, which will in general be lower than the 
	$\sqrt{s_{0}}$ of the incident nucleons. From this I can deduce an 
	attenuation factor $X$ in different ways:
	\begin{enumerate}
			\item  For high $p_{T}$ production it is assumed that the 
			multiplicities scale according to 
			$(1-x_{T})^{9}$ \cite{prl:antreasyan:77} with $x_{T} \equiv 
			2p_{T} / \sqrt{s}$, 
			so one gets:
			\begin{equation}
				X_{high} = \frac{\left(1-2p_{T}/\sqrt{s_{1}}\right)^{9}}
				{\left(1-2p_{T}/\sqrt{s_{0}}\right)^{9}}.
				\label{eq:xdec1}
			\end{equation}

			\item  For low $p_{T}$ production a scaling with $\ln(s)$ is 
			assumed \cite{npb:thome:77}:
			\begin{equation}
				X_{low} = \frac{0.88 + 0.44 \ln(s_{1}) 
				+ 0.118 \ln^{2}(s_{1})}
				{0.88 + 0.44 \ln(s_{0}) + 0.118 \ln^{2}(s_{0})}.
				\label{eq:xdec2}
			\end{equation}

			\item  The most radical assumption is that of complete 
			absorption. A nucleon produces particles only in the very first 
			collision:
			\begin{equation}
				X_{abs} = \left\{ \begin{array}{r@{,\qquad}l}
					1 & N_{pre}=0, \\
					0 & N_{pre}>0.
				\end{array}
				\right.
				\label{eq:xdec3}
			\end{equation}
			This corresponds to the limit of maximum momentum transfer in 
			cases 1) and 2).
			
	\end{enumerate}
	These factors are averaged over all possible binary collisions and 
	attenuation factors $x_{i} = \left\langle X_{i} \right\rangle$ are obtained.
	
	\item The 
	\emph{effective number of prescatterings} $N_{eff}$. For this, the 
	squareroot of the number of prescatterings in each individual binary 
	collision is calculated. This is then averaged over all possible 
	binary collisions. It should 
	control the effects of initial state multiple scattering. The value of 
	$N_{eff}$ for a specific reaction depends on the way the averaging 
	over the different binary collisions is done. For each binary 
	collision the corresponding attenuation factor $X_{i}$ has to 
	be taken into account as a weight, because it reduces the 
	multiplicity. 
	This introduces an implicit dependence on the longitudinal rapidity 
	shift $\delta y$ -- a larger deceleration will result in a smaller 
	effective number of prescatterings.
	
	A certain transverse rapidity shift 
	$\delta \rho$ is attributed to every prescattering -- this is the shift 
	apparent in the final particle production, so the rapidity shift a 
	single nucleon gets is actually $2 \delta \rho$. 
	Since $N_{eff}$ already includes 
	the effects of random relative orientation of these prescatterings 
	by adding the contributions quadratically, the average transverse 
	rapidity shift in a given nuclear reaction is just 
	$\rho(N_{eff}) = \delta \rho \cdot N_{eff}$. 
	The effect of the additional transverse velocity is then calculated 
	with the help of a formula frequently used to describe hydrodynamical 
	transverse flow 
	\cite{prc:sollfrank:hydro:91,prc:schnedermann:hydro:93}:
	\begin{equation}
		\frac{1/p_{T} \cdot dN/dp_{T}}{N_{coll}} = C \cdot m_{T} \cdot 
		\mathrm{I}_{0}\left(\frac{p_{T} \,
		\mathrm{sinh}\left[\rho(N_{eff})\right]}{T}\right)
		\mathrm{K}_{1}\left(\frac{m_{T} \, 
		\mathrm{cosh}\left[\rho(N_{eff})\right]}{T}\right).
		\label{eq:ptkick2}
	\end{equation}

\end{itemize}

With $N_{i}$ being the number of pre-scatterings for a given binary 
collision, the two parameters explained above are different averages
$N_{pre} = \left\langle N_{i} \right\rangle_{\left| i \right.}$ and 
$N_{eff} = \left\langle \sqrt{N_{i}} \right\rangle_{\left| i 
\right.}$.\footnote{One should note, however, that while $N_{eff}$ 
contains an influence from energy degradation there is no such effect 
included in $N_{pre}$.} 
They both have a well defined role within the model 
studied here, however, they also provide in general 
(esp. in the case $\delta y = 0$) measures of the effective thickness 
of the two colliding nuclei. It is therefore illustrative to study 
particle production as a function of these variables.

\begin{figure}[p]
	\centerline{\includegraphics{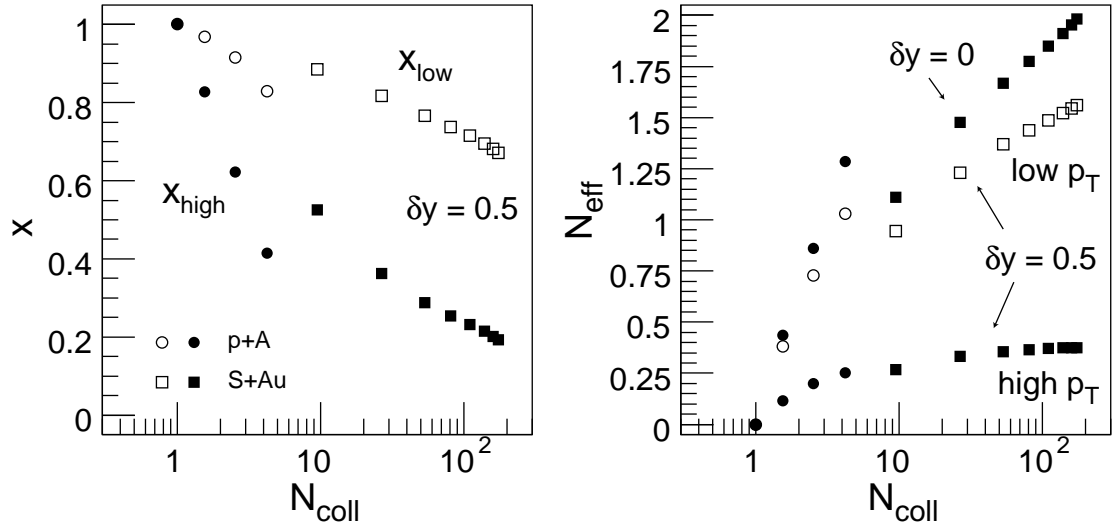}}
	\caption{Parameter values obtained in the model calculation for p+A  
	and centrality selected S+Au collisions. Left: attenuation factors 
	$x_{i}$
	for low- and high-$p_{T}$ production assuming $\delta y = 0.5$, 
	right: effective number of prescatterings $N_{eff}$ calculated 
	without deceleration ($\delta y = 0$) and 
	for low- and high-$p_{T}$ production assuming $\delta y = 0.5$. The 
	filled symbols show results relevant for high $p_{T}$, the open 
	symbols for low $p_{T}$.}
	\protect\label{fig:glauber1}
\end{figure}

Figure~\ref{fig:glauber1} shows sample results of this calculation for 
p+A and S+Au collisions at 200$A$GeV. 
The attenuation factor (left side of figure~\ref{fig:glauber1}) 
must obviously be $x=1$ 
for p+p collisions (i.e. $N_{coll} = 1$). It decreases for larger 
system size. There is however no unique dependence on $N_{coll}$, the 
behaviour is slightly different for S+Au compared to p+A. One also 
notices immediately that the effects are much stronger for high 
$p_{T}$ than for low $p_{T}$ -- this is a direct consequence of the 
stronger $\sqrt{s}$ dependence of the high 
$p_{T}$ production.

The effective number of prescatterings (right side of figure~\ref{fig:glauber1}) 
is also by definition $N_{eff} = 0$ for p+p collisions and increases 
with system size. There is a similar discontinuity when going from p+A 
to S+Au collisions. The largest values for $N_{eff}$ are obtained 
without any deceleration effect ($\delta y = 0$), a finite rapidity 
loss per collision of $\delta y = 0.5$ reduces these numbers slightly 
for low $p_{T}$ and drastically for high $p_{T}$.

\section{Analysis and Discussion}
The pion cross sections for $p_{T} = 0.77 \, \mathrm{GeV}/c$ 
and $p_{T} = 3.08 \, \mathrm{GeV}/c$
from \cite{prd:antreasyan:79} have been 
normalized with the nuclear inelastic cross section for the 
corresponding nuclei \cite{PDG} to obtain multiplicities/event. Neutral pion 
equivalent multiplicities 
have been calculated by averaging the results for negative and 
positive pions.

The neutral pion cross sections for S+Au
can be very well described by a fit with a power-law like formula 
\cite{misc:albrecht:pi0:98,phd:stueken}:
\begin{equation}
	\label{eqn:hagedorn}
	E\frac{d^3\sigma}{dp^3} = C^{\prime} \left( \frac{p_0}{p_T+p_0} \right)^n
\end{equation}
From this parameterization the pion multiplicities have been 
calculated at the same $p_{T}$ values as above. The systematic 
uncertainty of the pion yield given in \cite{misc:albrecht:pi0:98} 
will be used as error. 

The reaction cross sections from \cite{misc:albrecht:pi0:98} for the 
different centrality classes selected by the transverse energy $E_{T}$ are used 
to calculate impact parameter estimates assuming a monotonic increase 
of $E_{T}$ with $b$. This may lead to relatively large uncertainties 
for the most peripheral reactions which are affected most strongly by 
the experimental trigger bias. While the results presented below 
include an estimate of the systematic error of the model predictions 
from variations of the model assumptions, this may not completely 
account for the trigger bias in the most peripheral trigger class.

Scaled pion multiplicities are then calculated for different input 
parameters of the model:
\begin{equation}
	R_{\pi}
	= \frac{1}{N_{coll}} 
	\cdot E\frac{d^3N}{dp^3}\left( p_{T} \right)
	\cdot \frac{1}{X_{i}(\delta y)}.
	\label{eq:scale1}
\end{equation}

\begin{figure}[p]
	\centerline{\includegraphics{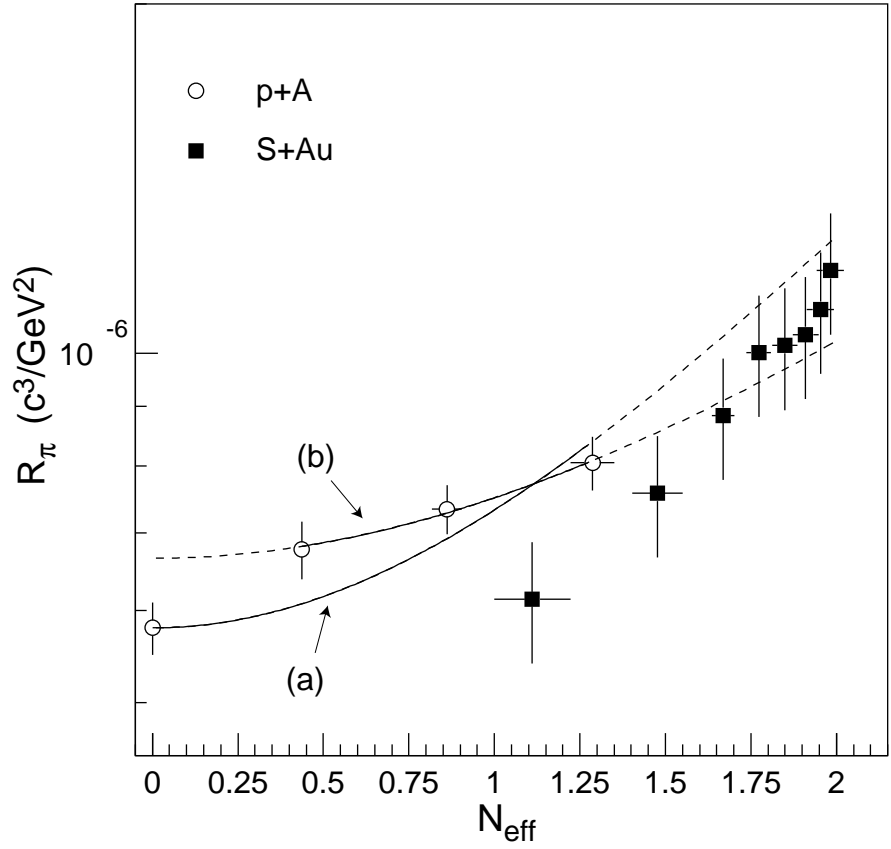}}
	\caption{Normalized particle multiplicities at
	$p_{T} = 3.08 \, \mathrm{GeV}/c$ for p+A and S+Au 
	collisions as a function of the effective number of prescatterings 
	$N_{eff}$. The lines show fits of the multiple scattering 
	model (equation~\ref{eq:ptkick2}) to the p+A data including (a) and 
	excluding (b) the p+p data point.}
	\protect\label{fig:ptkick1}
\end{figure}

Figure~\ref{fig:ptkick1} displays the scaled multiplicities at
$p_{T} = 3.08 \, \mathrm{GeV}/c$ for p+A and 
S+Au as a function of the effective number of prescatterings $N_{eff}$. 
The upper left part shows the data without any correction for 
deceleration effects. One immediately realizes that the multiplicity per 
collision rises with $N_{eff}$ (i.e. with the target mass) 
in p+A collisions (open circles) -- this is just a different 
representation of the anomalous nuclear enhancement.

The p+A data have been fitted with equation \ref{eq:ptkick2} 
according to the following procedure: The slope parameter $T$ was 
tuned to describe p+p at high $p_{T}$ -- this yields 
$T = 240 \, \mathrm{MeV}$. With $T$ fixed the other parameters were 
then fitted to the data for the heavier targets in two different 
ways: a) keeping the normalization fixed to describe p+p and b) 
leaving the normalization as a free parameter. Both curves are shown 
in figure~\ref{fig:ptkick1}. The fit (a) does not correctly describe 
all the p+A data, while fit (b) accounts well for the heavier targets 
but misses the p+p point. The same analysis can be done assuming a 
finite amount of deceleration of the participant nucleons, the 
qualitative picture of the dependence being unaltered.
%
The fit results are summarized in table~\ref{tbl:ptkick}. It can be 
seen that the necessary transverse rapidity shift $\delta \rho$ increases 
with larger (longitudinal) rapidity loss $\delta y$ of the participants. One 
should note that for $\delta y = 1$ a transverse rapidity 
shift of $\delta \rho \approx 7$ would be necessary to account for the 
rise in the data. This is completely unphysical. 

\begin{figure}[tb]
	\centerline{\includegraphics{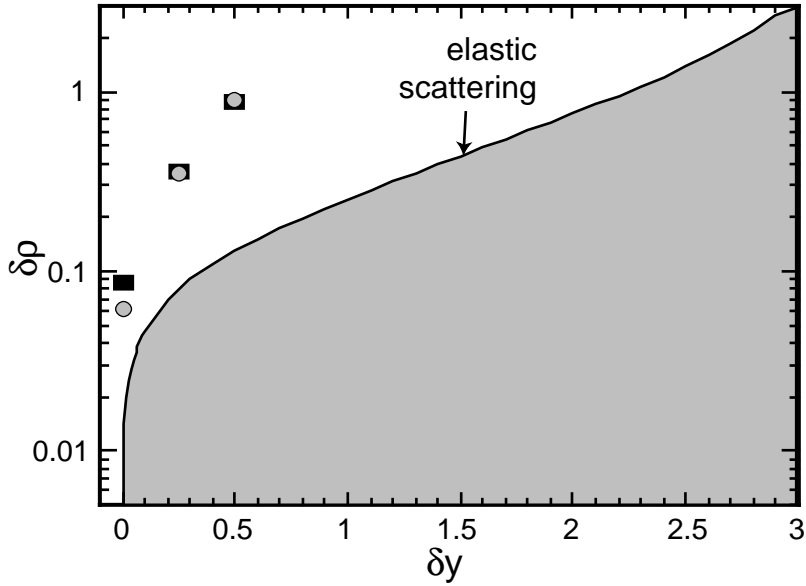}}
	\caption{
	Transverse rapidity $\delta \rho$ as a function of the 
	longitudinal rapidity loss $\delta y$ for nucleon-nucleon scattering 
	processes at $\sqrt{s} = 19.4 \, \mathrm{GeV}$. 
	The solid line shows the relation for elastic scattering 
	(equation~\ref{eq:rapid}), 
	the symbols show the fit parameters of the multiple scattering model 
	given in table \ref{tbl:ptkick}, where the black squares make use of 
	the p+p data point and the grey circles are obtained without it.}
	\protect\label{fig:rapid}
\end{figure}

In this context it is of interest to compare the parameters 
necessary to account for the enhancement with the kinematics of 
scattering processes. 
Every gain in net transverse momentum $p_{T}^{\prime}$ is related to a 
loss in momentum $p_{L}^{\prime}$ available for particle production. 
For elastic scattering one can easily deduce from momentum 
conservation a 
relation between the transverse rapidity shift $\delta \rho_{elast}$ 
a nucleon obtains in a collision and the 
longitudinal rapidity loss $\delta y_{elast}$ it suffers:
\begin{equation}
	\left[ \tanh(y_{0}) \right]^{2} =
	\left[ \tanh(y_{0} - \delta y_{elast}) \right]^{2} +
	\left[ \tanh(\delta \rho_{elast}) \right]^{2}.
	\label{eq:rapid}
\end{equation}
This relation is displayed in figure~\ref{fig:rapid} as a solid line. 
For any inelastic process, the transverse rapidity can only be 
smaller than this limit for a given longitudinal rapidity loss. This 
allowed region is indicated in figure~\ref{fig:rapid} as a grey area.
As a good approximation this relation can also be used as a limit for 
the transverse shift obtained in a multiple scattering 
picture.\footnote{The estimate would be different for cases, 
where the collision axis of the binary collision is strongly tilted 
relative to the overall collision axis - this will not be considered 
here.}
The symbols show the parameter values extracted from the fits of 
equation~\ref{eq:ptkick2} to the data. It is obvious that these values 
are not consistent with scattering kinematics. The maximum transverse 
rapidity shift allowed for a given $\delta y$ would not suffice to 
explain the enhancement observed in p+A collisions.

This observation can most easily be interpreted as a failure of the 
underlying picture of incoherent scatterings. This is actually 
expected because of the short time scales involved in 
hard scattering processes. In a coherent model there is no possibility 
of an energy degradation between two ``individual'' collisions like it 
is implemented in the model used here, and thus the kinematical 
considerations do not apply in the same way. The results obtained may 
be regarded as a warning against too na{\"{\i}}ve use of such 
simplified models.
I will therefore not 
discuss these cases further. 

The S+Au data have also been included in figure~\ref{fig:ptkick1}. 
For all cases the data agree with the extrapolation from p+A for 
intermediate centralities, however, the increase with $N_{eff}$ is 
much stronger than predicted by the extrapolation. The difference in 
system size between p+A and S+Au data may be relevant: While the 
parameter $N_{eff}$ is comparable e.g. for p+W and peripheral S+Au, 
the number of binary collisions in peripheral S+Au is larger by a factor 
of two. 

Ignoring the data points for the peripheral S+Au collisions 
the heavy ion data 
would be consistent with the heavier target p+A data. 
In this case, however, the p+p data cannot be described. There is no 
universal  quantitative description of both p+A and S+Au data within 
this multiple scattering picture, although it seems to be able to predict 
the scaling behaviour for both groups of data qualitatively. 

\medskip

\begin{figure}[p]
	\centerline{\includegraphics{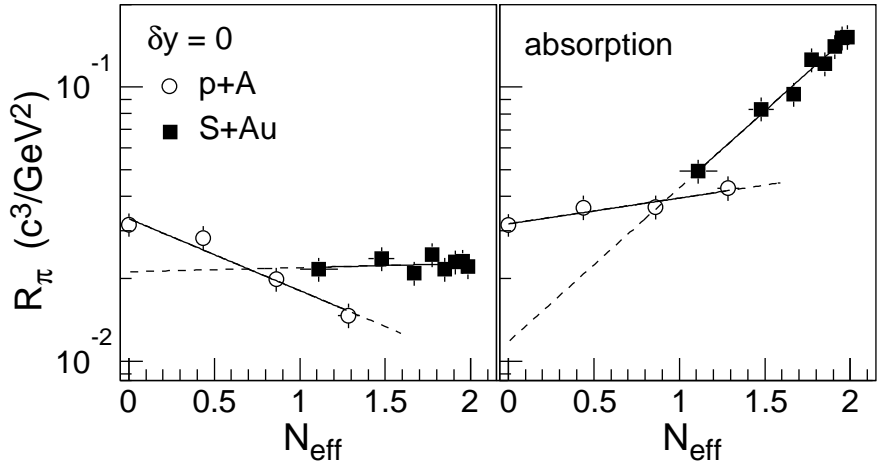}}
	\caption{Normalized particle multiplicities at
	$p_{T} = 0.77 \, \mathrm{GeV}/c$ for p+A and S+Au 
	collisions as a function of the effective number of prescatterings 
	$N_{eff}$ assuming no rapidity loss (left) and complete absorption 
	(right). The lines show exponential fits to guide the eye. 
	In the representation of the data in the right plot the same 
	values as for $\delta y = 0$ have been used for $N_{eff}$, because the 
	true values would be $N_{eff} \equiv 0$.
	}
	\protect\label{fig:ptkick2}
\end{figure}

\medskip
Still, as calculations using a ``random-walk'' picture have been 
advocated for the description of nucleus-nucleus collisions 
\cite{qm96:leonidov}, 
I will compare the present model to particle production at 
lower transverse momentum ($p_{T} = 0.77 \, \mathrm{GeV}/c$). 
Figure~\ref{fig:ptkick2} shows the scaled multiplicities for this case. 
Without any deceleration effects (upper left) the multiplicity per 
collision decreases for increasing target size in p+A reactions. Like 
for the high $p_{T}$ case this reflects the value of the 
``Cronin''-exponent $\alpha$ -- here one finds $\alpha < 1$, 
which indicates that the nuclei are not at all transparent for these 
softer processes.

Introducing a correction for the attenuation with intermediate values 
of $\delta y = 0.25-0.5$ does not really compensate for the decrease. 
If one assumes complete ``absorption'', so 
that only the first binary collision contributes to particle 
production, the scaled multiplicities show a small 
increase.\footnote{A rapidity loss $\delta y \geq 2$ is necessary to compensate 
for the decrease. This is for practical purposes very similar to 
the complete absorption case.}
The particle yields at low $p_{T}$ in p+A can obviously be understood without 
any strong enhancement via initial state multiple scattering. They 
are very close to the expectation from production only at the surface 
of the nuclei. The 
multiplicities in S+Au collisions show an additional increase 
compared to the expectation from p+A.

\section{Conclusions}

Pion production at high $p_{T}$ scales approximately with the number of binary 
collisions.
The enhancement relative to this scaling in p+A and S+Au has been studied as a 
function of an effective thickness parameter. Results indicate that 
it will be difficult to describe the enhancement in both data sets 
with a unique function of the thickness.

The description of the nuclear enhancement at high $p_{T}$ 
in a picture of incoherent 
scatterings (``random-walk'') 
in
the present calculation 
would require parameters which 
contradict momentum conservation in individual scatterings. This 
confirms that the $p_{T}$-broadening at high $p_{T}$ from initial state 
multiple scattering should not be described incoherently.  
Recent perturbative QCD
calculations including the effects of intrinsic $p_T$ and initial
state multiple scattering \cite{wang:1998:qcd} were able to 
reproduce data for neutral pion production from central 
S+S reactions \cite{misc:albrecht:pi0:98} 
as well as preliminary 
results for central Pb+Pb collisions 
\cite{Peitzmann:1996:qm96,Peitzmann:1997:qm97}.
It would be of interest to see, whether a QCD calculation 
with coherent multiple scattering effects
would be able to describe also the scaling of particle multiplicity 
with centrality
in heavy ion collisions.

Particle production at intermediate $p_{T}$ in p+A is consistent with 
production only at the nuclear surface. 
In this scenario there is little room for an enhancement by
initial state multiple scattering. It must therefore be considered 
very unlikely that e.g. the increase in inverse slope of the momentum 
spectra with system size also at intermediate $p_{T}$ can 
be attributed to this mechanism in the way discussed in \cite{qm96:leonidov}.

A simple explanation of increasing slopes with larger system 
size might be an admixture of the two components, where the soft 
component dominates at low $p_{T}$ with increasing importance of the hard 
component with increasing $p_{T}$. The ratio hard/soft would roughly 
scale as volume/surface and would thus increase the high $p_{T}$ part 
of the spectra more rapidly with increasing system size. This could 
explain the behaviour in p+A collisions at intermediate $p_{T}$. The 
particle production in S+Au collisions might call for additional 
thermal or hydrodynamical production not present in p+A, 
while for the explanation of the high $p_{T}$ production more refined 
calculations of hard scattering appear to be necessary.
%
\begin{ack}
	The author would like to thank D. St{\"u}ken for discussions 
	regarding the WA80 data and for providing the fits to the data. 
	Enlightening discussions with X.-N. Wang are also acknowledged.
\end{ack}
%
%

%
\begin{table}[p]
	\begin{center}
	\caption{Transverse rapidity shift per collision $\delta \rho$ from a fit 
	of equation \ref{eq:ptkick2} to data at $p_{T} = 3.08 \, 
	\mathrm{GeV}/c$ for p+p and p+A reactions at 200 GeV 
	\cite{prd:antreasyan:79}.}
		\begin{tabular}{|c|c|c|}
			\hline
			 & $\delta \rho$ (p+p and p+A) & $\delta \rho$ (only p+A)  \\
			\hline \hline
			$\delta y = 0$ & $0.087 \pm 0.006$ & $0.061 \pm 0.015$  \\
			\hline
			$\delta y = 0.25$ & $0.357 \pm 0.011$ & $0.348 \pm 0.029$  \\
			\hline
			$\delta y = 0.5$ & $0.882 \pm 0.022$ & $0.903 \pm 0.073$  \\
			\hline
			$\delta y = 1$ & $6.8 \pm 0.2$ & $7.1 \pm 0.5$  \\
			\hline
		\end{tabular}
	\label{tbl:ptkick}
	\end{center}
\end{table}
%
\end{document}